\documentclass[fleqn,12pt,twoside]{article}
\usepackage{espcrc1}

\usepackage{graphicx}
\usepackage[figuresright]{rotating}

\newcommand {\sNN}     {\mbox{$\sqrt{s_{NN}}$}}
\newcommand {\kT}      {\mbox{$k_{T}$}}
\newcommand {\mkT}     {\mbox{$\langle k_{T}\rangle$}}
\newcommand {\Rs}      {\mbox{$R_{side}$}}
\newcommand {\Ro}      {\mbox{$R_{out}$}}
\newcommand {\Rl}      {\mbox{$R_{long}$}}
\newcommand {\Ros}     {\mbox{$R_{out}/R_{side}$}}

\title{
       Two-particle correlations measured by PHENIX in Au-Au collisions
       at $\sNN=200$ GeV.
      }

\author{
        A. Enokizono\address[HIRO]{Hiroshima University,
        1-3-1 Kagamiyama, Higashi-Hiroshima 739-8526, Japan}
        for the PHENIX Collaboration\thanks{for the full PHENIX
        Collaboration author list and acknowledgements, see Appendix
        "Collaborations" of this volume.}
       }

\begin{document}

\maketitle

\begin{abstract}
In 2001, PHENIX measured particle correlations of identical
charged pions and kaons at mid-rapidity in Au-Au collisions at \sNN~=
200 GeV.
Bertsch-Pratt radius parameters of pion pairs are studied for 9
\kT~regions from 0.2 to 2.0 GeV/c and for 9 selections of collision
centralities. The pion radius parameters are consistent with the result
at \sNN~= 130 GeV, and the ratio \Ros~is below 1 in all \mkT~ranges up
to 1.2 GeV/c. The radius parameters from charged kaon correlations are
compared with those of pions.
\end{abstract}

\section{Introduction}
Particle interferometry provides a tool to study the final stage of
relativistic heavy ion collisions. Intensity interferometry is based
on the Hanbury-Brown Twiss, or HBT, effect.
The HBT analysis in relativistic high-energy heavy-ion experiments was
originally motivated by theoretical predictions that a large source
size and the long duration of particle emission \cite{GYULAP} would be
observed due to
softening of the equation-of-state in a first-order phase transition
from a quark-gluon-plasma state.
To analyze the HBT data, the Bertsch-Pratt parameterization is employed
in a longitudinal co-moving (LCMS) frame, where the three-dimensional
Gaussian radii are parameterized to be \Rs, \Ro~and \Rl~\cite{PrattP}.
Assuming a cylindrically symmetric, longitudinally expanding,
transversely homogeneous source, \Rs~corresponds to a radial size of
the source, while  \Ro~is a combination of the radial size and a
temporal term of the source duration time.
To study the space-time evolution of the source, we introduce two
independent external parameters: the transverse momentum of the
particle pair and the collision centrality. 
The transverse momentum, \kT, of the pair is defined by
$k_{T}=(p_{T1}+p_{T2})/2$, where $p_{Ti}$ is the transverse momentum of
each particle in the pair.
Recently, transport models \cite{HeinzP,HiranoP} have been used to
predict the radius parameters and their \kT~dependence. We will
compare our results to these predictions.

\begin{figure}[htb]
  \begin{minipage}[h]{0.5\textwidth}
    \includegraphics[width=8cm,height=6.5cm]{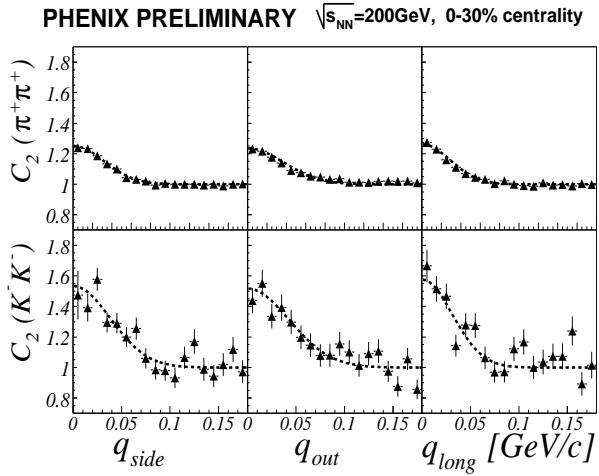}
  \end{minipage}
  \hfill
  \parbox[h]{0.45\textwidth}{
    \caption{Three-dimensional correlation functions for Bertsch-Pratt
    radius parameters for $\pi^{+}$ pairs (top) and $K^{-}$ pairs
    (bottom). The projection of 3-D correlation functions are averaged
    over the lowest 40 MeV in the orthogonal directions. The error
    bars are statistical only. The lines correspond to fits over the
    entire distribution. \kT~range is from 0.2 to 2.0 GeV/c with the
    \mkT~being 0.46 GeV/c for $\pi^{+}$ and 0.76 GeV/c for
    $K^{-}$. \label{fig:kaon3D}}
  }
\end{figure}

\section{Analysis}
The PHENIX detector provides powerful particle identification (PID)
capabilities over a wide momentum range \cite{JeffP}.
In this analysis, charged pions and kaons are identified by the
time-of-flight technique using the beam-beam counters and the
electromagnetic calorimeter (EMCal), combined with momentum
measurement by the drift chamber (DCH).
The detector covers the pseudorapidity region $|\eta|<0.35$ and 
$\Delta\phi=\pi/2$ in azimuthal angle.
The timing resolution is approximately 500 ps.
Charged pions and kaons are identified in squared mass versus momentum
space using momentum-dependent PID bands which take into account the
EMCal timing resolution and DCH momentum resolution.
Charged pions and kaons must lie within
1.5$\sigma_{m^{2}}$ of their squared mass peak, but 2.5$\sigma_{m^{2}}$
away from neighboring PID bands.
Using 50 million minimum bias events, $\sim$70 million
charged pions and $\sim$6 million charged kaons are selected in a
momentum range from 0.2 to 2.0 GeV/c. 
Similar to our previous analyses at \sNN~= 130 GeV \cite{PHENIX130P},
two-track separation cuts were applied using the DCH and EMCal such
that pairs within 1 cm in the longitudinal direction and 0.02 radians
in the azimuthal direction of each other are removed to eliminate
ghost tracks and DCH inefficiency. 
Pairs that share the same cluster in the EMCal are also removed. 
The systematic error of the resultant radius parameters from these pair
cuts is about 10$\%$ in total. The systematic error is evaluated by
varying the condition of pair cuts, and the amount of the changes of
radius parameters.
After the pair cuts, about 160 million charged pion pairs and 1
million charged kaon pairs remain.
The statistics are 70 times larger than that obtained in the previous
runs \cite{PHENIX130P}.
A full Coulomb correction was applied assuming a Gaussian source
without pairs coming from resonance decays \cite{PrattP2}. 
The systematic error originating from the Coulomb correction is small
compared to the systematic errors for the pair cuts.
Figure~\ref{fig:kaon3D} shows projections of the three-dimensional
correlation function onto $q_{side}$, $q_{out}$ and $q_{long}$ for
positive pions (top) and negative kaons (bottom) after the full
Coulomb correction.

\begin{figure}[htb]
  \begin{minipage}[h]{0.5\textwidth}
    \includegraphics[width=9.8cm,height=8.0cm]{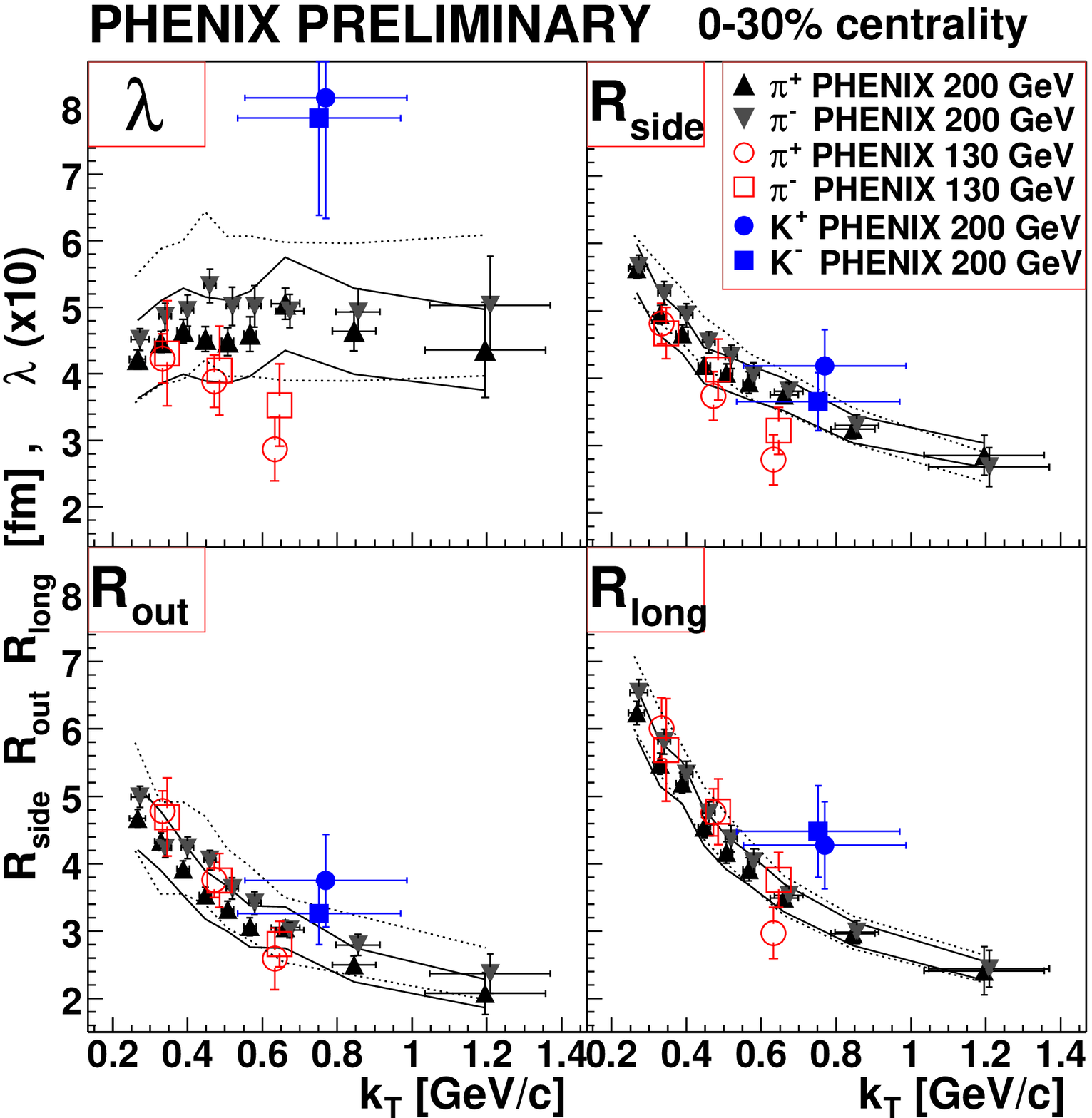}
  \end{minipage}
  \parbox[h]{0.33\textwidth}{
    \caption{The \kT~dependence of Bertsch-Pratt radius parameters
    and $\lambda$ for $\pi^{\pm}$ with statistical error bars and
    systematic error bands. The horizontal error bars indicate the
    root-mean-square of the \kT~distribution.
    PHENIX results of $\pi^{\pm}$ at 130 GeV and $K^{\pm}$ at 200 GeV
    are shown with statistical error bars.
    \label{fig:kTdep}}
  }
  \begin{minipage}[h]{0.65\textwidth}
    \includegraphics[width=9cm,height=6.5cm]{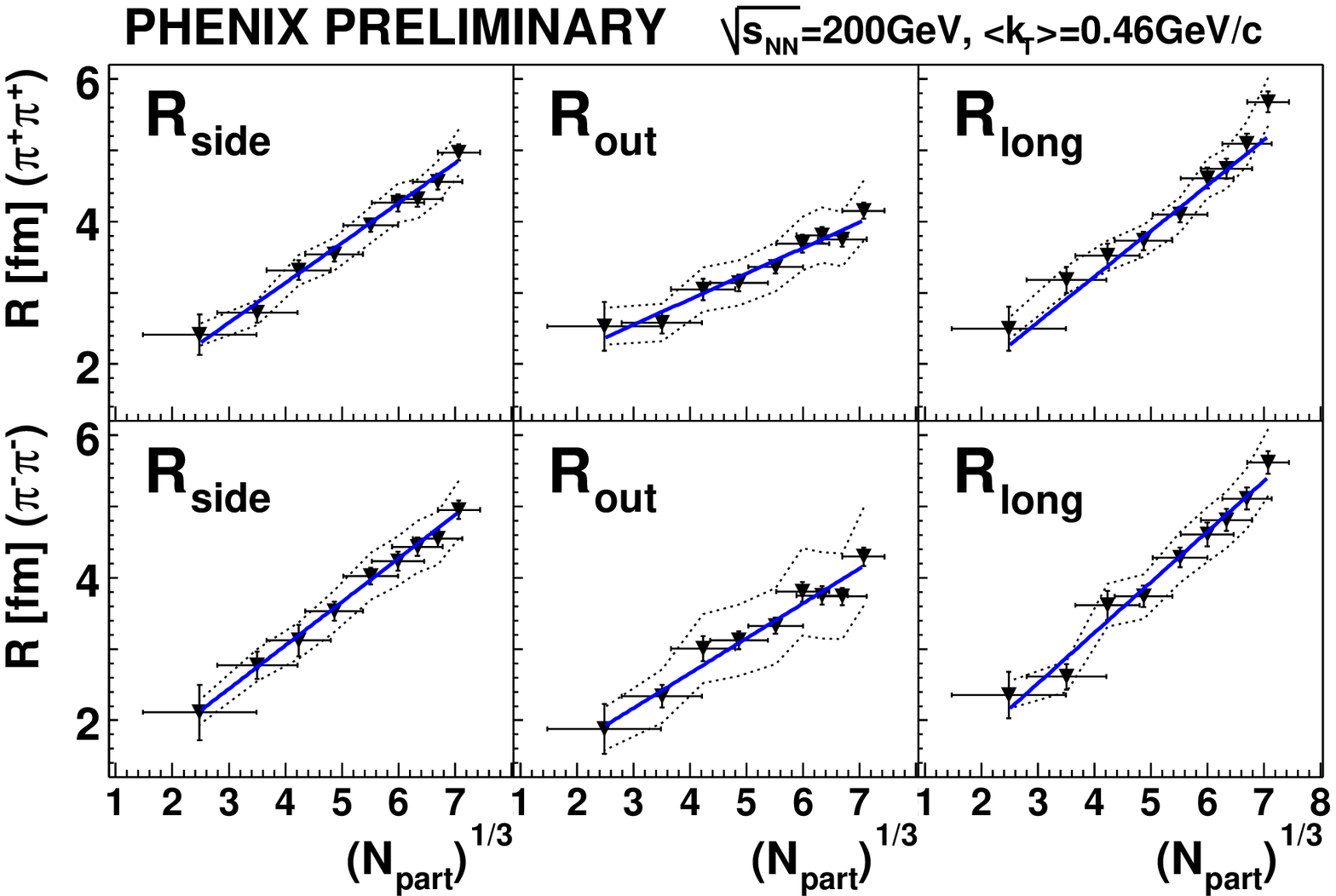}
  \end{minipage}
  \hfill
  \parbox[h]{0.35\textwidth}{
    \caption{Bertsch-Pratt radius parameters versus the number of participants
    ($N_{part}^{1/3}$) for $\pi^{\pm}$ with statistical error bars and systematic
    error bands. \kT~range is from 0.2 to 2.0 GeV/c with the
    \mkT~being 0.46 GeV/c. The solid lines show fits with a linear
    function of $N_{part}^{1/3}$. \label{fig:Npdep}}
  }
\end{figure}

\begin{figure}[htb]
  \begin{center}
    \includegraphics[width=16cm,height=7cm]{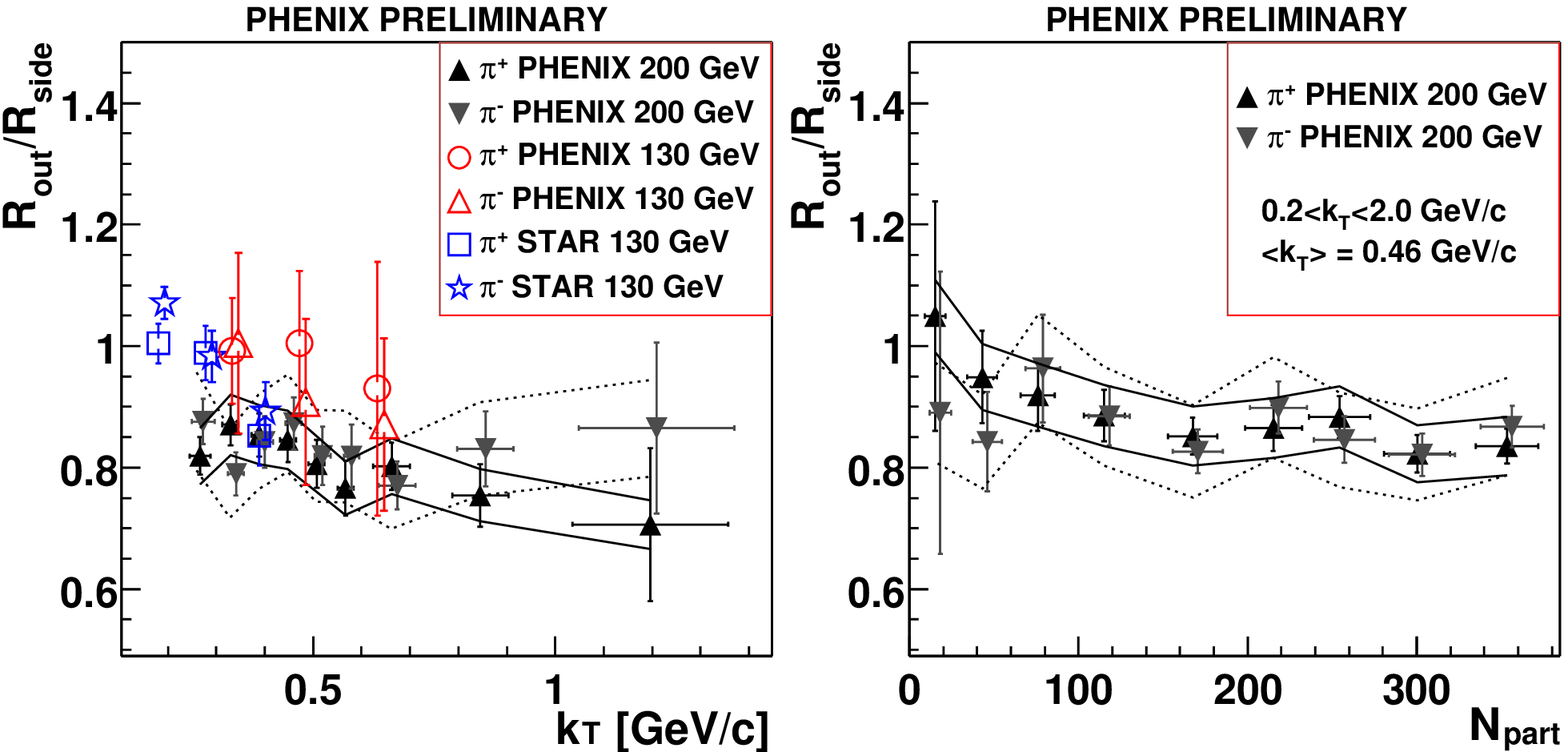}
    \caption{The left (right) figure shows the ratio \Ros~as a
    function of \kT~($N_{part}$) with statistical error bars and
    systematic error bands. In the left figure, results from PHENIX
    and STAR at 130 GeV \cite{PHENIX130P,STAR130P} are shown. The
    error bars are statistical only.
    \label{fig:Routside}}
  \end{center}
\end{figure}

\section{Results}

Figure \ref{fig:kTdep} shows the \kT~dependence of the chaoticity
and the Bertsch-Pratt radius parameters of pions at the most central
30$\%$ of the cross section.
The radius parameters at \sNN~= 200 GeV are, within errors, equal to
the corresponding results at 130 GeV \cite{PHENIX130P} at the same
\mkT.
While the radius parameters show a strong \kT~dependence, the
chaoticity is constant in all the \kT~bins.
The kaon radius parameters are slightly larger than
those of pions at the same \mkT~(the \kT~bin for kaons is between 0.2
to 2.0 GeV/c) as shown in Fig.~\ref{fig:kTdep}; however, the kaon
chaoticity is significantly larger than that of pions.
Figure \ref{fig:Npdep} shows the collision centrality dependence of
the radius parameters. The number of participants ($N_{part}$) is
evaluated from the charged particle multiplicity using a Glauber
model calculation \cite{PHENIX130M}.
The radius parameters depend linearly on $N_{part}^{1/3}$.
In order to study the duration time, we plot \Ros~in Fig. 
\ref{fig:Routside} as a function of (a) \kT~and (b) centrality.
The ratio is independent of centrality and is constant for \kT~up to
1.2 GeV/c within error, and is systematically below one.

\section{Summary}
We have presented HBT results in the Bertsch-Pratt frame for
identified charged pions and kaons measured by PHENIX in Au+Au
collisions at \sNN~= 200 GeV.
We have obtained the three-dimensional correlation functions with 160
million charged pion pairs and 1 million kaon pairs.
A clear \kT~dependence of the pion radius parameter was found, in
agreement with the results at \sNN~= 130 GeV.
The kaon radius parameters are equal to or slightly larger
than the pion radius parameters at the same \mkT~bin.
The pion radius parameters are consistent with a linear
parameterization as a function of $N_{part}^{1/3}$.
It should be noted that the measured radius parameters in the
transverse directions (\Rs~and \Ro) contradict recent transport model
predictions \cite{HeinzP,HiranoP}. However, the measured longitudinal
radius parameter (\Rl) is well reproduced by those models.
The \Ros~ratios are independent of both \kT~and centrality, and
are systematically smaller than unity albeit with large errors.
No evidence is seen for the anomalous increase of \Ros~as predicted
from calculations based on the formation of a QGP \cite{BassP}.

\end{document}